\let\ifarx=\iftrue

\ifarx
\documentclass[10pt,a4paper,final]{article}
\usepackage[backref,
            natbib
           ,style = numeric-comp
           ,maxnames = 2
           ,backend = bibtex
           ,sorting=none
           ]{biblatex}
\else
\documentclass[
           aps,
           twocolumn,
           prl,
           showkeys,
           showpacs,
           preprintnumbers,
           nofootinbib,
           final
]{revtex4-1}
\fi
  
\usepackage[T1]{fontenc} % Use 8-bit encoding that has 256 glyphs
\usepackage[utf8]{inputenc} % http://inspirehep.net/info/faq/general#utf8 
\ifarx \usepackage{palatino,eulervm} \fi
\usepackage[english]{babel} % English language/hyphenation
\usepackage{csquotes}
\usepackage{amsmath,amssymb,amsfonts,amsthm} % Math packages
\usepackage[table,dvipsnames*,svgnames]{xcolor}
\usepackage{fixltx2e}
\usepackage{showkeys}
\usepackage{nicefrac}
\usepackage{slashed}

\ifarx \usepackage[top=2.5cm, bottom=2.5cm, outer=3cm, inner=3cm, heightrounded, marginparwidth=2cm, marginparsep=0.5cm]{geometry} \fi
\usepackage[active]{srcltx}
\usepackage{extarrows}

\usepackage{braket}
\usepackage{mathrsfs}

\ifarx
\makeatletter

\let\@authors\@empty
\let\@email\@empty
\let\@affiliationone\@empty
\let\@affiliationtwo\@empty
\let\@pdfsubject\@empty
\let\@keywords\@empty
\let\@preprint\@empty

\providecommand{\pdfsubject}[1]{\gdef\@pdfsubject{#1}}
\providecommand{\keywords}[1]{\gdef\@keywords{#1}}
\renewcommand{\author}[1]{\ifx\@authors\@empty\toks@\expandafter{#1}\else\toks@\expandafter{\@authors, #1}\fi\edef\@authors{\the\toks@}}
\providecommand{\email}[1]{\ifx\@email\@empty\toks@\expandafter{#1}\else\toks@\expandafter{\@email, #1}\fi\edef\@email{\the\toks@}}
\providecommand{\affiliationone}[1]{\gdef\@affiliationone{#1}}
\providecommand{\affiliationtwo}[1]{\gdef\@affiliationtwo{#1}}
\providecommand{\preprint}[1]{\gdef\@preprint{#1}}

\makeatother

\fi

\usepackage[hyperindex,breaklinks]{hyperref}

\hypersetup{
    unicode=false,          % non-Latin characters in Acrobat’s bookmarks
    pdftoolbar=true,        % show Acrobat’s toolbar?
    pdfmenubar=true,        % show Acrobat’s menu?
    pdffitwindow=false,     % window fit to page when opened
    pdfstartview={FitH},    % fits the width of the page to the window
    pdfproducer={Brown University}, % producer of the document
    pdfnewwindow=true,      % links in new window
    colorlinks=true,       % false: boxed links; true: colored links
    linkcolor=black,          % color of internal links (change box color with linkbordercolor)
    citecolor=DarkGreen,        % color of links to bibliography
    filecolor=magenta,      % color of file links
    urlcolor=cyan!30!black!70         % color of external links
}

\usepackage{graphicx}

\allowdisplaybreaks[3]
\newcommand{\del}{\partial}

\newcommand{\bal}{\begin{align}}
\newcommand{\eal}{\end{align}}
\newcommand{\scri}{\mathscr{I}}

\renewcommand{\[}{\begin{equation}}
\renewcommand{\]}{\end{equation}}
\ifarx
\bibliography{sbms.bib}
\fi
\begin{document}
\ifarx
\title{Residual Local Supersymmetry and the Soft Gravitino}

\author{Steven G.\ Avery\textsuperscript{a}}
 \email{\href{mailto:steven\_avery@brown.edu}{steven\_avery@brown.edu}}

\affiliationone{%
 \textsuperscript{a}Brown University\\ Department of Physics\\ 182 Hope St, Providence, RI 02912}

\author{Burkhard U.\ W.\ Schwab\textsuperscript{b}}
 \email{\href{mailto:burkhard\_schwab@brown.edu}{burkhard\_schwab@brown.edu}}

\affiliationtwo{%
\textsuperscript{b}Harvard University\\Center for Mathematical Science and Applications\\1 Oxford St, Cambridge, MA 02138}

\date{\today}
\keywords{Supergravity, soft limits, supersymmetry, BMS transformations}
\pdfsubject{Avery, Schwab - Residual Local Supersymmetry and the Soft Gravitino}
\preprint{Brown-HET-1689}

%%%%%%%%%%%%%%%%%%%%%%%%%%%
% Title Page

\makeatletter
\thispagestyle{empty}

\begin{flushright}
\begingroup\ttfamily\@preprint\par\endgroup
\end{flushright}

\begin{centering}
\begingroup\Large\normalfont\bfseries\@title\par\endgroup
\vspace{1cm}

\begingroup\@authors\par\endgroup
\vspace{5mm}

\begingroup\itshape\@affiliationone\par\endgroup
\vspace{3mm}
\begingroup\itshape\@affiliationtwo\par\endgroup
\vspace{3mm}

\begingroup\ttfamily\@email\par\endgroup
\vspace{0.25cm}

\begin{minipage}{13cm}
 \begin{abstract}
We show that there exists an infinite tower of fermionic symmetries in pure $d=4$, $\mathcal{N}=1$ supergravity on an asymptotically flat background. The Ward identities associated with these symmetries are equivalent to the soft limit of the gravitino and to the statement of supersymmetry at every angle. Additionally, we show that these charges commute into charges associated with the (unextended) BMS group, providing a supersymmetrization of the BMS translations.
 \end{abstract}
\end{minipage}

\vspace{3mm}
\rule{\textwidth}{.5mm}
\vspace{-1cm}

\end{centering}

\makeatother

%\tableofcontents

\else
\title{Residual Local Supersymmetry and the Soft Gravitino}
\date{\today}
\author{Steven G.\ Avery}
\email{steven_avery@brown.edu}
\affiliation{Brown University, Dept of Physics, 182 Hope St, Providence, RI 02912}
\author{Burkhard U.\ W.\ Schwab}
\email{schwa@cmsa.fas.harvard.edu}
\affiliation{Harvard University, CMSA, 1 Oxford St, Cambridge, MA 02138} 

\begin{abstract}
We show that there exists an infinite tower of fermionic symmetries in pure $d=4$, $\mathcal{N}=1$ supergravity on an asymptotically flat background. The Ward identities associated with these symmetries are equivalent to the soft limit of the gravitino and to the statement of supersymmetry at every angle. Additionally, we show that these charges commute into charges associated with the (unextended) BMS group, providing a supersymmetrization of the BMS translations.
\end{abstract}

\pacs{11.10.Jj,11.30.Pb,04.65.+e}
\keywords{Supergravity, soft limits, supersymmetry, BMS transformations}
\preprint{Brown-HET-1689}

\maketitle
\fi

\section{Introduction}
\label{sec:introduction}

Thanks to the remarkable series of papers~\cite{Strominger:2013jfa, Strominger:2013lka, Strominger:2014pwa, Strominger:2015bla, He:2014cra, He:2014laa, He:2015zea, Kapec:2014zla, Kapec:2015ena, Kapec:2015vwa, Lysov:2014csa, Banks:2014iha, Banks:2015pfi, Campiglia:2014yka, Campiglia:2015kxa, Campiglia:2015lxa, Campiglia:2015qka, Campiglia:2015yka} there has been tremendous progress in understanding the physical impact of \emph{large} or \emph{residual} gauge freedom not only in the semiclassical treatment of gravity, but also in Yang--Mills theory, Maxwell theory, string theory~\cite{Avery:2015gxa,Schwab:2014sla,Schwab:2014fia,Bianchi:2014gla,Volovich:2015yoa,DiVecchia:2015oba}, and most recently $\mathcal{N}=1$ SQED~\cite{Dumitrescu:2015fej}.

Here we provide an understanding of residual local supersymmetry of the Rarita--Schwinger (RS) field in the simplest setting, $d=4$, $\mathcal{N}=1$ supergravity. Remarkably, many of our results have been anticipated in \cite{Grisaru:1977kk}. However, we use the methods introduced in~\cite{Avery:2015rga} to connect the soft gravitinos of \cite{Grisaru:1977kk} with the asymptotic Killing spinors found in \cite{Awada:1985by}. The previously neglected asymptotic Killing spinors that asymptote to angle dependent spinors at asymptotic null infinity $\scri$ are contained in local supersymmetry transformations. They supersymmetrize the supertranslations of the BMS group and their Ward identities generate the soft limit of the gravitino which may then be understood as a statement of \emph{supersymmetry at every angle} in the language of~\cite{Strominger:2013jfa}.

\ifarx We first introduce the setting of $d=4$, $\mathcal{N}=1$ supergravity and derive the Noether charge density associated with local supersymmetry transformations from the action. Subsequently we show that there are residual gauge symmetries at $\scri$, calculate the associated algebra of charges, and relate the Ward identity to the soft limit of gravitinos.\fi

While preparing this article, we became aware of the independent work of Vyacheslav Lysov~\cite{lysov-new} on this topic, which supports our conclusions.

\section{Supergravity}
\label{sec:supergravity}

The $d=4$, $\mathcal{N}=1$ supergravity action in the 1.5 order formalism proves most convenient for our problem. It is $S = S_{2} + S_{\nicefrac{3}{2}}$ where
\begin{align}
  \label{eq:1}
  S_{2} &= \frac{1}{2\kappa^{2}}\int d^{4}x e\ e^{\mu a}e^{\nu b} R(\omega)_{\mu\nu ab}\\
  S_{\nicefrac{3}{2}} &= \frac{i}{2\kappa^{2}}\int d^{4}x \epsilon^{\mu\nu\rho\sigma}\overline{\psi}_{\mu}\gamma_{5}\gamma_{\sigma}D_{\nu}\psi_{\rho}.
\end{align}
The notation requires some explanation. Given a metric $g_{\mu\nu}$, define the vielbein field $e^{a}_{\mu}$ via  \[g_{\mu\nu}=e^{a}_{\mu}e^{b}_{\nu}\eta_{ab}\] with $\eta = {\rm diag}(-1,+1,+1,+1)$ and $e = \det e^{a}_{\mu}=\sqrt{-g}$. The $\gamma_{\mu}$-matrices are given by the contraction of the numerical matrices $\gamma_{a}$ in a given Lorentz frame with the frame field $e^{a}_{\mu}$. The covariant derivative $\nabla_{\mu}$ is given by 
\[
\nabla_{\mu}\psi_{\nu} = D_{\mu}\psi_{\nu} + \Gamma^{\kappa}_{\mu\nu}\psi_{\kappa},
\] 
where $D_{\mu}\psi_{\nu} = \del_{\mu}\psi_{\nu} + \omega_{\mu ab}\gamma^{ab}\psi_{\nu}$ is the spin covariant derivative. We used the explicitly four dimensional form of $S_{\nicefrac{3}{2}}$ with $\gamma_{5} = i\gamma_{0}\gamma_{1}\gamma_{2}\gamma_{3}$, the product of the numerical matrices. Note that, in the 1.5 order formalism, the spin connection $\omega_{\mu ab}$ is an a priori independent variable. By solving its (algebraic) equation of motion, one finds
\[\omega_{\mu ab} = \omega_{\mu ab}(e) + K_{\mu ab},\] where the first part is defined by a function of the frame field and its derivatives. The contortion $K_{\mu ab}$ is quadratic in the gravitino field $\psi_{\mu}$. Their exact forms can be found in, e.g., \cite{freedman-proeyen}, and are inessential for the current discussion. The curvature two-form is defined via
\[[D_{\mu},D_{\nu}] = \frac{1}{8}R_{\mu\nu a b}\gamma^{ab}.\]

\section{Noether two-form}
\label{sec:noether-two-form}

For simplicity we assume a purely bosonic asymptotically flat background. Thus any terms proportional to the torsion $T_{\mu\nu}{}^{a}$ in the following will be dropped. The transformations that define local supersymmetry are given by the gauge transformation of the RS field and a corresponding transformation on the frame field (which we could drop in our chosen background)
\begin{align}
  \label{eq:2}
  \delta e^{a}_{\mu}= \frac12 \overline{\epsilon}\gamma^{a} \psi_{\mu},\qquad  \delta \psi_{\mu}= D_{\mu}\epsilon
\end{align}
where $\epsilon$ is an \emph{anticommuting} spinor.

It was shown~\cite{Barnich:2001jy,Avery:2015rga} (see also \cite{Iyer:1994ys}) that there is a Noether charge density two-form $k^{\mu\nu}$ that is associated with every local symmetry. To derive $k$ for the transformations in \eqref{eq:2}, we use the formalism developed in~\cite{Avery:2015rga}. 

Any variation of the action may be written as \[\delta S = \int d^{4}x (-E^{I}\delta\phi_{I} + \del_{\mu}\theta^{\mu}(\phi_{I},\delta \phi_{I}))\] where $E^{I}$ are the equations of motion for the set of fields $\phi_{I} = \{e_{\mu}^{a},\psi_{\mu}\}$. For supersymmetry variations \eqref{eq:2}, $\theta^{\mu}$ is given by
\[\theta^{\nu} = \frac{1}{2\kappa^{2}} \left(i\epsilon^{\mu\nu\rho\sigma}\overline{D_{\rho}\epsilon}\gamma_{5}\gamma_{\sigma}\psi_{\mu}\right).\] Conversely, the action is only symmetric under \eqref{eq:2} up to a total derivative term \[\delta S = \frac{1}{2\kappa^{2}} \int d^{4}x \del_{\nu}\left(i\epsilon^{\mu\nu\rho\sigma}\overline{\psi}_{\rho}\gamma_{5}\gamma_{\sigma}D_{\mu}\epsilon\right) = \int d^{4} x \del_{\mu}K^{\mu},\] where we dropped terms in $\theta$ and $K$ proportional to the variation of the spin connection $\delta \omega_{\mu}$ since they will not contribute to the result. These two total derivative contributions define the ordinary Noether current $j^{\mu} = \theta^{\mu} - K^{\mu}$. In this combination the aforementioned terms proportional to $\delta\omega_{\mu}$ drop out. Finally, we need to find the weakly vanishing current $S^{\mu}$ from Noether's identity
\[E^{I}\delta\phi_{I} = \Delta_{I} E^{I} + \del_{\mu}S^{\mu}\] for local variations of the form $\delta\Phi_{I} = f_{I}(\Phi)\lambda + \sum_{i}f_{I}^{\mu_{1}\ldots\mu_{i}}\del_{\mu_{1}}\cdots\del_{\mu_{i}}\lambda$. For \eqref{eq:2} this is 
\[S^{\mu} = \frac{i}{\kappa^{2}}\left(\epsilon^{\mu\nu\rho\sigma}\overline{\epsilon}\gamma_{5}\gamma_{\sigma}D_{\nu}\psi_{\rho}\right)\] and so 
\[\del_{\mu}k^{\nu\mu} = j^{\nu}-S^{\nu} = \del_{\mu}\Big[ \frac{e}{\kappa^{2}}(\overline{\epsilon}\gamma^{\nu\mu\kappa}\psi_{\kappa})\Big]\label{eq:twoform}\] where for the last form we made use of \eqref{eq:threetofive}. Although we didn't make this explicit in the calculation, it is possible to show that the torsion does not enter $k$.

\section{Boundary conditions and Asymptotic Killing Spinors}\label{sec:aks}

Following the approach of~\cite{Avery:2015rga}, we must gauge fix the local, bounded supersymmetry transformations to define the path integral measure. We use Witten gauge $\ifarx\slashed{\psi}\else \gamma\cdot\psi\fi = 0$.
In Witten gauge, the gravitino wave equation becomes $\slashed{D}\psi_\nu = 0$ with leading asymptotic behavior in Bondi coordinates~\eqref{eq:bondi},
\begin{equation}\label{eq:psi-asymp}
\psi_u = \psi^{(0)}_u(\theta) \qquad
\psi_r = \psi^{(0)}_r(\theta) \qquad
\psi_A = r\,\psi^{(0)}_A(\theta).
\end{equation}
The equation of motion fixes the subleading behavior.

The two-form takes the form
\begin{equation}
k^{\mu\nu} = \frac{1}{2\kappa^2} \bar{\epsilon}
   \big(\gamma^{\nu\mu}\ifarx\slashed{\psi}\else\gamma\cdot\psi\fi + 2\gamma^{[\mu}\psi^{\nu]}\big).
\end{equation}
Demanding the charge is finite for the asymptotic Killing spinors in Witten gauge implies the fall-off condition
\begin{equation}\label{eq:fall-off}
2\gamma^{[u}\psi^{r]} = O\left(r^{-2}\right).
\end{equation}
The initial data in~\eqref{eq:psi-asymp} is constrained by the gauge condition $\ifarx\slashed{\psi}\else\gamma\cdot\psi\fi = 0$, and the fall-off condition~\eqref{eq:fall-off}. To wit, we may take the spinors on the sphere $\psi^{(0)}_A$ as the physical data.

The gauge fixing condition leaves unfixed a discrete set of large residual supersymmetries, or asymptotic Killing spinors. These have already been discussed from a different perspective~\cite{Awada:1985by}: there is an infinite family, which are the ``square root'' of the BMS supertranslations.

The residual supersymmetries are parametrized by spinors solving the Dirac equation, $\slashed{D}\epsilon = 0$,  for which the transformation~\eqref{eq:2} preserves the boundary conditions discussed above. Solutions are parametrized by spinors that asymptote to arbitrary angle-dependent spinors $\eta(\theta)$ on $\scri^\pm$,
\begin{equation}\label{eq:killspinors}
\epsilon = \eta(\theta) + O\left(r^{-1}\right).
\end{equation}
The ``small'' subleading pieces are gauge-dependent and do not contribute to the charges.

\section{Algebra}
\label{sec:algebra}

Now, define a charge $Q[\eta]$ for large supersymmetry transformations $\eta$. $Q[\eta]$ can be written as \[Q[\eta] = \int_{\sigma}\star k = \frac{1}{\kappa^{2}}\int_{\sigma} dS_{\mu\nu}\,\overline{\eta}\gamma^{\mu\nu\rho}\psi_{\rho}\label{eq:charge}\] where $\sigma\subset\scri$ is an $S^{2}$ at $u=-\infty$ on $\scri^{+}$. Then~\cite{Hull:1983ap, Witten:1981mf, Nester:1982tr} \[\Big[Q[\eta_{1}],Q[\eta_{2}]\Big] = \delta_{\eta_{1}}Q[\eta_{2}] = \frac{1}{\kappa^{2}}\int_{\sigma} dS_{\mu\nu}\,\overline{\eta}_{2}\gamma^{\mu\nu\kappa}D_{\kappa}\eta_{1}\] which can be reformulated as 
\[\delta_{\eta_{1}}Q[\eta_{2}] =\frac{1}{2\kappa^{2}}\int_{\sigma}dS_{\mu\nu}\,\left(\overline{\eta}_{2}\gamma^{\mu\nu\kappa}D_{\kappa}\eta_{1}-\overline{(D_{\kappa}\eta_{2})}\gamma^{\mu\nu\kappa}\eta_{1}\right)\] where we made use of the compactness of $S^{2}$ to drop a total derivative term. The remainder takes the \emph{Nester--Witten} form of the diffeomorphism charge~\cite{Witten:1981mf, Nester:1982tr}. It is possible to rewrite this, see e.g., \cite{Cecotti:2015wqa}, using the linear approximation $D_{\mu} = \overline{D}_{\mu} + \Omega_{\mu} + O(h^{2})$ where $h$ is the perturbation to the background metric $g_{\mu\nu} - \overline{g}_{\mu\nu} = h_{\mu\nu}$, $\overline{D}$ the background derivative, and $\Omega$ the linear part. The frame field is then given by $e_{\mu}^{a} = \overline{e}_{\mu}^{a} + h_{\mu}^{a}$ such that $h_{\mu\nu} = e_{\mu}^{a}h_{\nu a} + e_{\nu}^{a}h_{\mu a}$. We use $\overline{g}$ to raise and lower coordinate indices and $\overline{e}_{\mu}^{a}$ to transform to a local Lorentz frame. Since $\Omega_{\mu} = \Omega_{\mu}{}^{\alpha\beta}\gamma_{\alpha\beta}$ and $\gamma_{\sigma}\gamma_{\alpha\beta} + \gamma_{\alpha\beta}\gamma_{\sigma} = 2 \gamma_{\sigma\alpha\beta}$ along with the identity \eqref{eq:threetofive} the commutator may be written as
\[\frac{1}{\kappa^{2}}\int_{\sigma} dS_{\mu\nu} \epsilon^{\mu\nu\kappa\sigma}\epsilon_{\sigma\alpha\beta\rho} \overline{\eta}_{2}\gamma^{\rho}\eta_{1} \Omega_{\kappa}{}^{\alpha\beta}.\] We dropped a term proportional to $\bar{\eta}_{[2}\gamma^{\mu\nu\kappa}\overline{D}_{\kappa}\eta_{1]}$ which may be though of as a potential central charge. Define now, as usual, $\xi^{\rho} = \overline{\eta}_{2}\gamma^{\rho}\eta_{1}$ as the parameter of a coordinate transformation and contract the antisymmetric tensors to get the result
\[\frac{6}{\kappa^{2}}\int_{\sigma} dS_{\mu\nu}\,\xi^{[\mu}\Omega_{\kappa}{}^{\nu\kappa]}.\] While it is almost a standard calculation, let us anyhow show that we can transform this result into the form of BMS charges. Note that the linearized spin connection $\Omega_{\kappa}{}^{\mu\nu}$ can be written as
\[\Omega_{\kappa}{}^{\mu\nu} = \overline{g}^{\sigma[\nu}\delta\Gamma^{\mu]}_{\sigma\kappa}\] with $\delta\Gamma^{\mu}_{\sigma\kappa} = \frac12 \overline{g}^{\mu\rho}(\overline{\nabla}_{\sigma}h_{\kappa\rho}+\overline{\nabla}_{\kappa}h_{\sigma\rho}-\overline{\nabla}_{\rho}h_{\kappa\sigma})$. Then $\Omega$ can be written in two ways
\begin{align}
  \label{eq:3}
  \Omega_{\kappa}{}^{\mu\nu} &= \frac12\overline{g}_{\kappa\alpha}\overline{\nabla}_{\sigma}\left(\overline{g}^{\sigma[\nu}h^{\mu]\alpha}\right)\notag\\
  &= \frac12\overline{g}_{\kappa\alpha}\overline{\nabla}_{\sigma}H^{\sigma\alpha\mu\nu} - \overline{\nabla}_{\sigma}\left(\delta^{[\mu}_{\kappa}h^{\nu]\sigma}-\delta^{[\mu}_{\kappa}\overline{g}^{\nu]\sigma}h\right)
\end{align}
where we defined the quantity $H^{\sigma\alpha\mu\nu} = \overline{g}^{\sigma\nu}\overline{h}^{\mu\alpha} + \overline{g}^{\mu\alpha}\overline{h}^{\sigma\nu} - \overline{g}^{\sigma\mu}\overline{h}^{\nu\alpha} - \overline{g}^{\alpha\nu}\overline{h}^{\mu\sigma}$, the trace $h = h^{\kappa}_{\kappa}$, and the trace reversed metric perturbation $\overline{h}^{\mu\nu} = h^{\mu\nu}-\frac12\overline{g}^{\mu\nu}h$. Inspecting the expression $\xi^{[\mu}\Omega_{\kappa}{}^{\nu\kappa]}$ and inserting the two expressions in \eqref{eq:3} for $\Omega$, we find that
\[\Big[Q[\eta_{1}],Q[\eta_{2}]\Big] = T[\xi]=\frac{1}{\kappa^{2}}\int_{\sigma} dS_{\mu\nu}\, \xi_{\alpha}\overline{\nabla}_{\sigma}H^{\sigma\alpha\mu\nu}\] which differs from the flux integral of the Barnich--Brandt Noether two-form~\cite{Barnich:2001jy} for diffeomorphisms by a boundary term and is thus equivalent under the integral. Note that $\xi^{\mu}$ as defined above using the asymptotic Killing spinors \eqref{eq:killspinors} has a finite value at the boundary, i.e., $T[\xi]$ constitutes a BMS translation~\cite{Bondi:1962px, Sachs:1962wk} in tune with \cite{Awada:1985by,Strominger:2013jfa,Grisaru:1977kk}. Furthermore, the bracket $[Q[\eta],T[\xi]] = 0$ as expected.

\section{Ward identity}
\label{sec:ward-identity}

The Ward identity associated with the two-form derived above is given by \cite{Avery:2015rga}
\[\langle\delta_{\eta}(\Phi_{1}\cdots\Phi_{n})\rangle = i\langle \Phi_{1}\cdots\Phi_{n}\left(\int_{\scri^{+}}\star j[\eta] - \int_{\scri^{-}}\star j[\eta]\right)\rangle \label{eq:ward}\] where $\Phi$ are the fields of $d=4$, $\mathcal{N}=1$ supergravity. Here $\delta\Phi$ are local supersymmetry transformations \eqref{eq:2} while the right hand side introduces the operator \eqref{eq:charge} into the path integral. Note that the Noether current $j[\eta]$ is related to the two-form $k$ in \eqref{eq:twoform} simply via $j^{\mu} = \del_{\nu}k^{\mu\nu}$ up to equations of motion.

Using Witten gauge $\gamma^{\mu}\psi_{\mu}=0$ and Bondi coordinates \eqref{eq:bondi}, the integral over $\scri^{+}$ on the right hand side of eq.~\eqref{eq:ward} can be written as (analogously for $\scri^{-}$)
\[Q[\eta]=\frac{1}{\kappa^{2}}\int_{\scri^{+}}d\Sigma_{\mu}\, \overline{\eta}\Big[\overleftarrow{D}_{\nu}\gamma^{\mu\nu\kappa}\psi_{\kappa} + \gamma^{\mu\nu\kappa}D_{\nu}\psi_{\kappa}\Big]\] similarly for $\scri^{-}$. All derivatives in this section ff.\ are background derivatives $\overline{D}$. We use the Witten gauge condition, the equations of motions $\gamma^{\mu\nu\rho}D_{\nu}\psi_{\rho} = \mathcal{J}^{\mu}$ with the supercurrent $\mathcal{J}^{\mu}$, and the Majorana flip to write
\[Q[\eta] = \frac{1}{\kappa^{2}}\int_{\scri^{+}}d\Sigma_{\mu}\,\Big[\overline{\eta}\mathcal{J}^{\mu} - \overline{\psi}^{\mu}\gamma^{\nu}D_{\nu}\eta + \overline{\psi}^{\nu}\gamma^{\mu}D_{\nu}\eta\Big].\] The second term vanishes due to the residual gauge condition $\gamma^{\mu}D_{\mu}\eta = 0$. Thus with $\eta|_{\scri}=\eta(z,\bar z)$
\[Q[\eta]=\frac{1}{\kappa^{2}}\int_{\scri^{+}} d^{2}z du\ \overline{\eta}\Big[\mathcal{J}^{r}+\overleftarrow{D}_{z}\gamma_{u}\psi_{\bar z} + \overleftarrow{D}_{\bar z}\gamma_{u}\psi_{z} \Big].\label{eq:softfact}\] 

\section{Gravitino Soft Factor}
\label{sec:soft-theorem}

From the Ward identities on correlation functions, we extend the result to $\mathcal{S}$ matrix elements by LSZ reduction following \cite{Grisaru:1976vm}. The boundary fields $\psi_{z,\bar{z}}|_{\scri}$ in \eqref{eq:softfact} can be found by a limiting process on the asymptotic mode expansions (see \eqref{eq:expansions}) as in \cite{Dumitrescu:2015fej,He:2014laa,He:2015zea,Kapec:2015ena,Kapec:2015vwa,He:2014cra,Kapec:2014zla,Lysov:2014csa}. The result is ($c^{\pm}_{p}$ is the gravitino annihilation operator)
\begin{align}
\psi_{z} = \frac{i\sqrt{2}}{8\pi^{2}(1+z\bar{z})}\int dE u^{+}_{E \hat{x}} (c^{+}_{E\hat{x}}e^{-iEu} - (c^{-}_{E\hat{x}})^{\dagger}e^{iEu})
\end{align} and $\psi_{\bar {z}}$ the same with interchanged helicities. We used $u^{\pm}=v^{\mp}$. The spinor, $u^{+}$  ($u^{-}$), is right-handed (left-handed) in a helicity basis for the $\gamma^{\mu}$. Thus we may use the projectors $P_{R}$ ($P_{L}$) to further reduce the charge to 
\[Q[\eta] = \frac{1}{\kappa^{2}}\int_{\scri^{+}}d^{2}z du\ \overline{\eta}\Big[\mathcal{J}^{r}+\overleftarrow{D}_{z}\gamma_{u}P_{L}\psi_{\bar z} + \overleftarrow{D}_{\bar z}\gamma_{u}P_{R}\psi_{z} \Big].\] Only $\psi_{z,\bar z}$ are dependent on the coordinate $u$, so we define $\Psi_{z,\bar z} = \lim_{E_{0}\to 0}\int du\,e^{iE_{0}u} \psi_{z,\bar z}$ with a regulating factor of $\exp(iE_{0}u)$ to extract the zero modes. $\eta$ can be chosen left-handed or right-handed to single out specific terms in $Q[\eta]$. The second and third term in the charge $Q[\eta]$ can even be localized if $\eta$ consists of a function $(z-z_{i})^{-1}$ (or $(\bar{z}-\bar{z}_{i})^{-1}$) multiplied by a right-handed (left-handed) spinor that is constant with respect to $D_{\bar z}$ (or $D_{z}$), i.e.,
\[\eta_{\dot\alpha} = \frac{1}{z-w}\chi_{\dot\alpha},{\rm\ or\ }\eta_{\alpha} = \frac{1}{\bar{z}-\bar{w}}\chi_{\alpha}.\label{eq:factors}\] More generally $\eta = \epsilon(z,\bar z) \chi$ with $\epsilon(z,\bar z)$ an arbitrary function. The zero mode of the supercurrent has trivial action on the particle vacuum, $\int du\mathcal{J}^{r}|0\rangle=0$, but the rest of the charge $Q[\eta]$ inserts a zero momentum gravitino and acts like \emph{soft charge} $Q_{s}[\eta]$ in the terminology of \cite{Dumitrescu:2015fej,He:2014laa,He:2015zea,Kapec:2015ena,Kapec:2015vwa,He:2014cra,Kapec:2014zla,Lysov:2014csa}, i.e.,  $Q[\eta] = Q_{h}[\eta] + Q_{s}[\eta]$ generates a spontaneously broken symmetry.

Finally, let us inspect the soft limit of scattering amplitudes $M_{n}$ with $n$ particles and one soft, positive helicity gravitino denoted $\psi^{+}_{s}$ with soft momentum $p_{s}$. Fermions come in pairs; there is at least one more gravitino in the set $\{1,\ldots, n\}$. All particles are considered outgoing. Then
\[\label{eq:softlim}
\lim_{p_{s}\to 0}M_{n+1}(\ldots,\psi_{s}^{+}) = \sum_{i=1}^{n}S^{\psi^{+}}_{i}M_{n}(\ldots,\mathcal{D}\Phi_{i},\ldots)
\]
where $S^{\psi^{+}}_{i} = \frac{\epsilon^{+}_{s}\cdot p_{i}}{p_{i}.p_{s}}\epsilon^{+}_{i,\mu}\bar{u}_{s}^{+}\gamma^{\mu}v_{i}^{-}$ and $\mathcal{D}$ lowers the helicity of the i\textsuperscript{th} external leg by $\nicefrac{1}{2}$
\[\mathcal{D}h^{+} = \psi^{+},\quad \mathcal{D}\psi^{-}=h^{-},\quad \mathcal{D}\psi^{+} = \mathcal{D}h^{-}=0.\] In the last two cases above, the amplitude on the right hand side vanishes. Using $p^{2}=0$, let $p_{\mu}\sigma^{\mu} = \lambda_{p}
\bar\lambda_{p}$ spinor helicity variables. Since gravitinos are Majorana we have $\bar{u}^{+}_{s} = (\bar{\lambda}_{s},0)$ and $v^{-}_{i} = (0,\lambda_{i})$ in a helicity basis. Then the polarization can be written as $\epsilon^{+}_{i}.\sigma = \frac{\lambda_{x}\bar{\lambda}_{i}}{\langle x,i\rangle}$ where $\lambda_{x}$ is an arbitrary reference spinor and $\sigma^{\mu}$ the Pauli matrices. In the last equation we employ the usual bracket notation, see, e.g., \cite{Elvang:2013cua}. Then $S^{\psi^{+}}_{i}$ may be written as $\frac{[s,i] \langle x,i\rangle}{\langle s,i\rangle \langle x,s\rangle}$, compare \cite{Liu:2014vva}. Thus the positive helicity gravitino soft limit is the combination of a helicity lowering supersymmetry transformation and the multiplication of an angle dependent factor $S^{\psi^{+}}_{i}$ \cite{Grisaru:1977kk}.

The soft limit \eqref{eq:softlim} can be related to the Ward identity \eqref{eq:ward} when employing an LSZ reduction \cite{Grisaru:1976vm}. Using asymptotic expansions for the graviton field $h_{\mu\nu}$ and the gravitino field $\psi_{\mu}$ (see \eqref{eq:expansions}) with in-state annihilators $a_{p}^{\pm}$, respectively $c_{p}^{\pm}$, for particles with chirality $\pm$, the action of a supersymmetry variation $\delta_{s}$ with large parameter $\eta$ is given by
\begin{align}
\delta_{\eta}a_{p}^{+} = [Q[\eta],a_{p}^{+}] = \epsilon_{p,\mu}^{+}\bar\eta \gamma^{\mu} v^{-}_{p}c_{p}^{+}\ifarx ,\\ \qquad \else \notag\\ \fi \delta_{\eta}c_{p}^{-} = [Q[\eta],c_{p}^{-}] = \epsilon_{p,\mu}^{+}\bar\eta \gamma^{\mu} v^{-}_{p} a_{p}^{-}
\end{align} where here, as above, $Q[\eta]$ bosonic. The other two cases are similar and would be necessary for a discussion of the Ward identity for a negative helicity gravitino.We match this onto the right hand side of the soft limit above by remarking that $S^{\psi^{+}}_{i}$ is a special case of the coefficient of the commutation relations above. This statement is clearer when writing $S^{\psi^{+}}_{i}$ with the help Bondi coordinates \[S^{\psi^{+}}_{i} = \frac{1+z_{s}\bar{z}_{s}}{\sqrt2 E_{s} (z_{s}-z_{i})}\bar{u}_{s}^{+}\epsilon^{+}_{i,\mu}\gamma^{\mu}v_{i}^{-}.\] We may discard the divergence in $E_{s}$ by multiplying \eqref{eq:softlim} with $\sqrt{E_{s}}$ since $\bar\lambda_{s}\propto \sqrt{E_{s}}$ as well as pull out the factor of $(1+z\bar{z})$. Then, if we let $\sqrt{E_{s}}^{\nicefrac{1}{2}}\eta = \frac{1}{z_{s}-z_{i}}u^{-}_{s}$ we see that it is exactly of the form given in \eqref{eq:factors}. It follows that the $\mathcal{S}$ matrix statement of the Ward identity \eqref{eq:ward} is the leading soft gravitino limit. The analogous statement for a negative helicity gravitino can be derived in the same way.

\ifarx \paragraph{Acknowledgments} \else \begin{acknowledgments} \fi
We are grateful to Vyacheslav Lysov for sharing a draft of his paper and coordinating the arXiv submission of our papers.
BS was supported by the Center for Mathematical Sciences and Applications at Harvard University and the Cheng Yu-Tung fund, and NSF grant 1205550. BS is thankful to Andrew Strominger and Thomas Dumitrescu for discussions.
SGA is supported by US DOE grant de-sc0010010.
\ifarx \else \end{acknowledgments}\fi

\appendix

\section{Conventions}
\label{sec:conventions}

We follow the conventions of~\cite{freedman-proeyen}. Throughout the paper, the preferred background is asymptotically flat space with metric $g^{\mu\nu}$, strictly bosonic, and excluding black hole spacetimes. Retarded Bondi coordinates are given by the metric
\begin{equation}\label{eq:bondi}
ds^2 = -du^2 - 2dudr + r^2 d\Omega^2
\end{equation}
with $d\Omega^2 = \frac{4}{(1+\vec{\theta}^2)^2}d\vec{\theta}^2 $. $\vec{\theta}\in \mathbb{R}^2$ covers the sphere once. In the text, we use $z = \theta_1 + i \theta_2$. The frame is defined by
\begin{equation}
e^0 = du + dr\qquad
e^1 = dr\qquad
e^A = \frac{2r}{1+\vec{\theta}^2}d\theta^A
\end{equation}
with spin connection,
\begin{equation}
\omega_{A1} = - \omega_{1A} = \frac{e^A}{r}\qquad
\omega_{AB} = \frac{1}{r}(\theta_A e_B - \theta_B e_A).
\end{equation}

The $\gamma$ Clifford algebra is defined in the usual way by $\{\gamma^{\mu},\gamma^{\nu}\} = 2 g^{\mu\nu}$. All calculations use the Majorana flip extensively; given two \emph{anticommuting} spinors $\xi$, $\chi$ and a set of $\gamma$-matrices, the flip is defined by
\[\overline{\xi}\gamma^{\mu_{1}}\cdots \gamma^{\mu_{r}}\chi = t_{r}\overline{\chi}\gamma^{\mu_{1}}\cdots\gamma^{\mu_{r}}\xi\] where a convenient choice for $t_{r}$ is $t_{r} = -1$ for $r=1,2{\rm\, mod\, } 4$ and $t_{r}= 1$ for $r=0,3{\rm\, mod\, } 4$ in $d=4$.

We also use the notation $\gamma^{\mu\nu\kappa} = \frac{1}{3!}\gamma^{[\mu}\gamma^{\nu}\gamma^{\kappa]}$ and the identity
\[\gamma^{\mu\nu\kappa} = -ie^{-1}\epsilon^{\mu\nu\kappa\sigma}\gamma_{5}\gamma_{\sigma}\label{eq:threetofive}.\]
Asymptotic expansions for the fields are given by
\ifarx \else \begin{widetext} \fi
\begin{align}\label{eq:expansions}
h_{\mu\nu}(x) &= \sum_{\sigma=\pm}\int\frac{d^{3}p}{(2\pi)^{3}2 E} (\epsilon^{\sigma,\star}_{p,\mu\nu}a^{\sigma}_{p}e^{ip.x} + \epsilon^{\sigma}_{p,\mu\nu}(a^{\sigma}_{p})^{\dagger}e^{-ip.x})\notag\\
\psi_{\mu}(x) &= \sum_{\sigma=\pm}\int\frac{d^{3}p}{(2\pi)^{3}2 E} (\epsilon^{\sigma,\star}_{p,\mu}u^{\sigma}_{p}c^{\sigma}_{p}e^{ip.x} + \epsilon^{\sigma}_{p,\mu}v_{p}^{\sigma}(c^{\sigma}_{p})^{\dagger}e^{-ip.x}).
\end{align}
\ifarx \else \end{widetext} \fi

\ifarx \printbibliography \else \bibliography{sbms} \fi

\end{document}